\documentclass[twocolumn]{aastex631}

\received{\today}
\revised{\today}
\accepted{\today}

\submitjournal{ApJ}

\shorttitle{EUI jet eruption}
\shortauthors{Long et al.}

\graphicspath{{./}{figures/}}

\newcommand{\corr}[1]{{\color{black}{{#1}}}} 

\begin{document}


\title{Multi-stage reconnection powering a solar coronal jet}
\correspondingauthor{David M.~Long}
\email{david.long@ucl.ac.uk}

\author[0000-0003-3137-0277]{David M. Long}
\affiliation{University College London, Mullard Space Science Laboratory, Holmbury St. Mary, Dorking, Surrey, RH5 6NT, UK}

\author[0000-0002-9270-6785]{Lakshmi Pradeep Chitta}
\affiliation{Max Planck Institute for Solar System Research, Justus-von-Liebig-Weg 3, 37077, Göttingen, Germany}

\author[0000-0002-0665-2355]{Deborah Baker}
\affiliation{University College London, Mullard Space Science Laboratory, Holmbury St. Mary, Dorking, Surrey, RH5 6NT, UK}

\author[0000-0003-1193-8603]{Iain G.~Hannah}
\affiliation{School of Physics \& Astronomy, University of Glasgow, University Avenue, Glasgow G12 8QQ, UK}

\author[0000-0002-1794-1427]{Nawin Ngampoopun}
\affiliation{University College London, Mullard Space Science Laboratory, Holmbury St. Mary, Dorking, Surrey, RH5 6NT, UK}

\author[0000-0003-4052-9462]{David Berghmans}
\affiliation{Solar-Terrestrial Centre of Excellence -- SIDC, Royal Observatory of Belgium, Ringlaan 3, 1180 Brussels, Belgium}

\author[0000-0002-2542-9810]{Andrei N. Zhukov}
\affiliation{Solar-Terrestrial Centre of Excellence -- SIDC, Royal Observatory of Belgium, Ringlaan 3, 1180 Brussels, Belgium}
\affiliation{Skobeltsyn Institute of Nuclear Physics, Moscow State University, 119992 Moscow, Russia}

\author[0000-0001-7298-2320]{Luca Teriaca}
\affiliation{Max Planck Institute for Solar System Research, Justus-von-Liebig-Weg 3, 37077, Göttingen, Germany}

\begin{abstract}

Coronal jets are short-lived eruptive features commonly observed in polar coronal holes and are thought to play a key role in the transfer of mass and energy into the solar corona. We describe unique contemporaneous observations of a coronal blowout jet seen by the Extreme Ultraviolet Imager onboard the \emph{Solar Orbiter} spacecraft (SO/EUI) and the Atmospheric Imaging Assembly onboard the \emph{Solar Dynamics Observatory} (SDO/AIA). The coronal jet erupted from the south polar coronal hole, and was observed with high spatial and temporal resolution by both instruments. \corr{This enabled identification of the different stages of a breakout reconnection process producing the observed jet.} We find bulk plasma flow kinematics of $\sim$100--200~km~s$^{-1}$ across the lifetime of its observed propagation, with a distinct kink in the jet where it impacted and was subsequently guided by a nearby polar plume. We also \corr{identify} a \corr{faint} faster feature ahead of the bulk plasma motion propagating with a velocity of $\sim$715~km~s$^{-1}$ which we attribute to untwisting of newly reconnected field lines during the eruption. A Differential Emission Measure (DEM) analysis using the SDO/AIA observations revealed a very weak jet signal, indicating that the erupting material was likely much cooler than the coronal passbands used to derive the DEM. This is consistent with the very bright appearance of the jet in the Lyman-$\alpha$ passband observed by SO/EUI. The DEM was used to estimate the radiative thermal energy of the source region of the coronal jet, finding a value of $\sim2\times10^{24}$~ergs, comparable to the energy of a nanoflare.

\end{abstract}


\section{Introduction} \label{sec:intro}

Coronal jets are collimated ejections of plasma from the solar atmosphere that could escape into the heliosphere. Typically observed using extreme ultraviolet (EUV) \citep[e.g.,][]{Alexander:1999,Nistico:2009,Chandra:2014a} or X-ray \citep[e.g.,][]{Shibata:1992,Shimojo:1996,Cirtain:2007,Sterling:2015} observations, they are better observed in coronal holes, due to the lower background intensity making them easier to identify \citep[e.g.,][]{Savcheva:2007}. They are the subject of detailed investigation, as they offer a mechanism for transferring mass and energy to the outer corona and solar wind \citep[cf.][]{Shen:2021}. They also represent an opportunity to study smaller scale evolution of the mechanisms involved in larger solar eruptions, with recent observations \citep[e.g.,][]{Nistico:2009,Sterling:2015} and simulations \citep[e.g.,][]{Wyper:2017,Wyper:2018} suggesting that coronal jets can be interpreted as being produced by the eruption of mini-filaments via breakout reconnection. A detailed overview of observations and modelling of coronal jets can be found in the recent reviews by \citet{Raouafi:2016} and \citet{Shen:2021}.

\begin{figure*}[!t]
    \centering
    \begin{interactive}{animation}{movie_0.mp4}
    \includegraphics[width=0.95\textwidth]{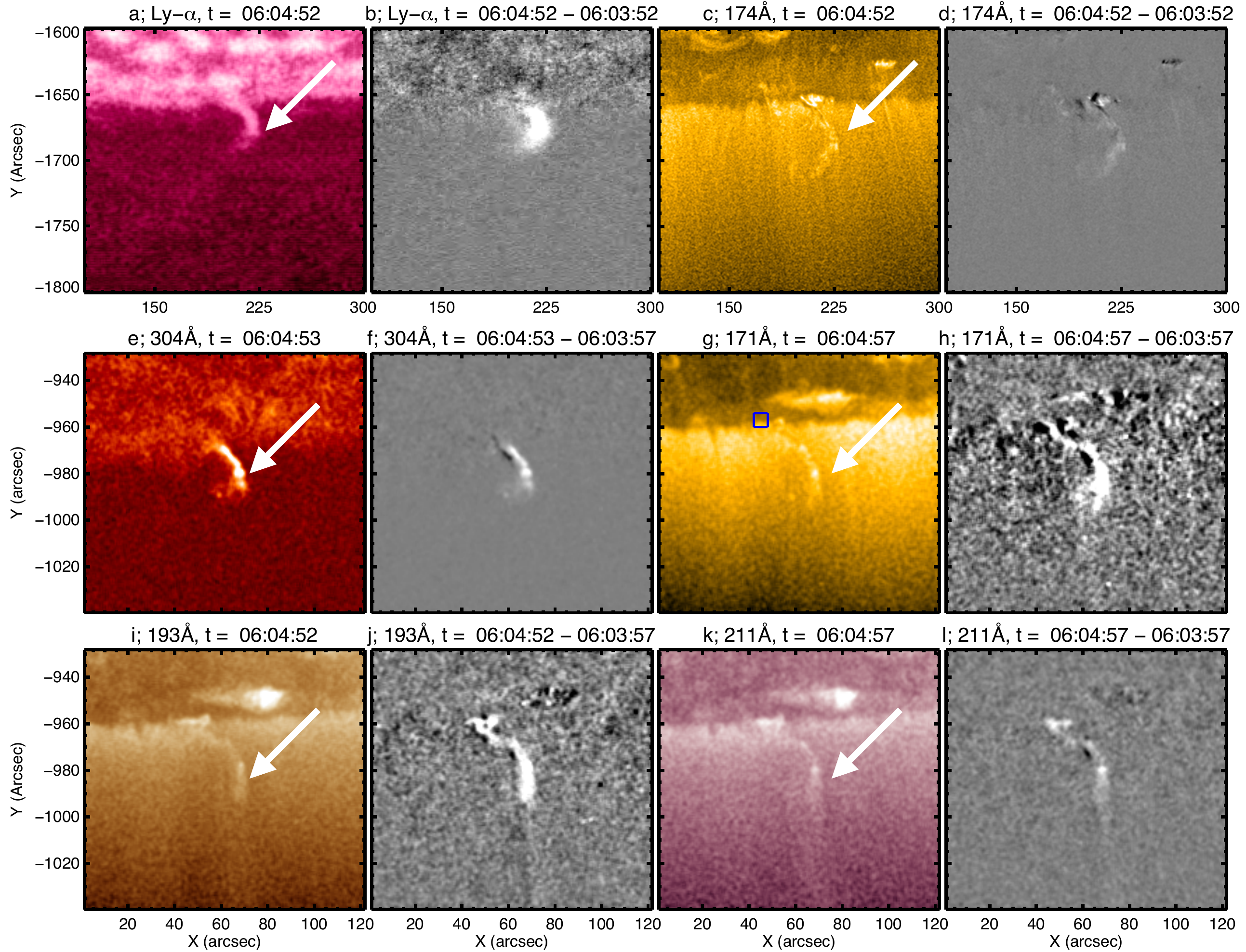}
    \end{interactive}
    \caption{Jet eruption in a polar coronal hole. The coronal jet (indicated by the white arrow) observed by EUI/HRI (top row) and SDO/AIA (middle and bottom rows) using a combination of intensity and running difference images for each passband. Top row shows the EUI/HRI 174~\AA\ and Lyman-alpha passbands, middle row shows the SDO/AIA 304~\AA\ and 171~\AA\ passbands, and the bottom row shows the SDO/AIA 193~\AA\ and 211~\AA\ passbands. In each case, intensity images have been enhanced using the Multiscale Gaussian Normalisation technique of \citet{Morgan:2014}, while the times of the images used to make the running difference images are given in the panel title. Note that all times indicate the time at Earth. An animated version of this figure is available as movie\_0.mp4, with a duration of 13~s which shows the temporal evolution of the erupting jet observed by both Solar Orbiter EUI and SDO/AIA.}
    \label{fig:context}
\end{figure*}

Coronal jets are typically identified as being divided into two distinct types, ``standard'' and ``blowout'' jets \citep[cf.][]{Moore:2010}, based on morphological behaviour. Standard jets follow the model originally proposed by \citet{Shibata:1992}, with a narrow spire that remains thin throughout the lifetime of the jet, a relatively dim source region, and no emission in the cooler 304~\AA\ passband. In contrast, blowout jets initially behave as standard jets, with the spire subsequently broadening to match the width of the footpoints. Blowout jets have also been observed to produce emission in the 304~\AA\ passband, with \citet{Moore:2010} suggesting that reconnection as a result of an emerging bipole which drives the jet could release the overlying magnetic field and enable the observed surge (i.e., the eruption of cooler material). This cooler material has traditionally been identified using the 304~\AA\ passband, but \citet{Alexander:1999} presented observations of a coronal jet made using the Lyman-$\alpha$ 1216~\AA\ passband from the \emph{Transition Region And Coronal Explorer} \citep[TRACE;][]{Handy:1999}. In this case, the jet produced bright emission in the Lyman-$\alpha$ passband, and could be tracked at a cadence of $\sim$60~s.

The multiple coronal extreme ultraviolet (EUV) passbands provided by the Atmospheric Imaging Assembly \citep[AIA;][]{Lemen:2012} onboard the \emph{Solar Dynamics Observatory} \citep[SDO;][]{Pesnell:2012} has enabled the development of multiple techniques to estimate plasma parameters such as density and temperature using differential emission measure \citep[DEM; see, e.g., the codes developed by][]{Aschwanden:2013,Hannah:2012,Hannah:2013,Plowman:2013,Cheung:2015,Pickering:2019}. This has enabled many plasma diagnostic studies of different solar phenomena, including coronal jets \citep[e.g.,][]{Chen:2013,Zhang:2014,Joshi:2020}. More recently, DEM analysis has also been used to probe the radiative thermal energy released during solar nanoflares, with \citet{Purkhart:2022} finding an energy range of $10^{24}$ -- $10^{29}$~erg for 30 SDO/AIA image series between 2011 and 2018. The  high spatial and temporal resolution provided by SDO/AIA has also enabled a detailed investigation of the evolution of coronal jets \citep[cf.][]{Morton:2012a,Morton:2012b,Chen:2012}. The High Resolution Imager (HRI) component of the Extreme Ultraviolet Imager \citep[EUI;][]{Rochus:2020} onboard the \emph{Solar Orbiter} \citep{Mueller:2020} spacecraft will enable much higher spatial and temporal resolution studies of solar phenomena in its 174~\AA\ and Lyman-$\alpha$ passbands. Already this is providing new insights into very small features associated with polar coronal jets \citep[e.g.,][]{Mandal:2022}, offering sub-arcsecond imaging of these features. 

In this work, we use a combination of observations from Solar Orbiter EUI and SDO/AIA to examine the initial evolution of a small scale coronal jet erupting from the southern polar coronal hole. The event is described in section~\ref{sec:obs}, with an analysis of the observations presented in section~\ref{sec:res} before some conclusions are drawn in section~\ref{sec:disc}.

\section{Observations and Data Analysis} \label{sec:obs}

The coronal jet discussed here erupted from the south pole of the Sun on 2021-Sept-14 beginning at $\sim$05:59:02~UT as observed by Solar Orbiter EUI\footnote{Note that the distances of the individual spacecraft from the Sun meant that phenomena were observed at different times by the different spacecraft, so to avoid confusion, we will use the time at Earth throughout this discussion.}. At the time, \emph{Solar Orbiter} was located at 0.587~astronomical units from the Sun at an angle of 47.372~degrees behind the Earth, and was doing a calibration manoeuvre, pointing at the North and South Poles, and East and West limbs to enable cross-calibration of the different high-resolution telescopes. The erupting jet was observed by both the 174~\AA\, and Lyman-$\alpha$ passbands with an image scale of 0.492\arcsec\ (1.028\arcsec) per pixel in the 174~\AA\ (Lyman-$\alpha$) passbands, and a temporal resolution of 5~s (see top row of Figure~\ref{fig:context}). For the analysis described here, we used the calibrated level-2 EUI/HRI data from EUI Data Release 4.\footnote{\url{https://doi.org/10.24414/s5da-7e78}}

The jet was also well observed by SDO/AIA, with the eruption starting at $\sim$06:02:30~UT as observed near Earth. SDO/AIA has an image scale of 0.6\arcsec\ per pixel at a 12~s cadence in each of the 7 EUV passbands (94, 131, 171, 193, 211, 304, 335~\AA), with the data presented here processed using the standard aia\_prep.pro routine contained within SolarSoftWare \citep{Freeland:1998}. The eruption was well observed by the 304, 171, and 193~\AA\ passbands (with their temperature response functions peaking at T$\sim10^{4.9}$, $10^{5.9}$, and $10^{6.2}$ respectively), and was also identifiable using the 211~\AA\ passband (T$\sim10^{6.25}$) as shown in the middle and bottom rows of Figure~\ref{fig:context}, with no signal apparent within the 94, 131, or 335~\AA\ passbands (T$\sim10^{6.85}$, $10^{5.7}$, and $10^{6.4}$ respectively). Note that a full description of the response function for each of the AIA passbands can be found in \citet{Boerner:2012}.

\section{Results} \label{sec:res}

The erupting jet is shown as observed using multiple passbands at $\sim$06:04:52~UT in Figure~\ref{fig:context}, with the temporal evolution of the jet shown in more detail in the associated animation movie\_0.mp4. The jet has a clear ``$\lambda$'' style morphology \citep[see e.g.,][for details]{Raouafi:2016}, with each passband also showing an apparent kink in the jet following its initial eruption as indicated by the white arrows. Inspection of the temporal evolution of the jet eruption using the larger field of view provided by SDO/AIA suggests that this kinking is due to the erupting jet interacting with a nearby very faint polar plume, which deflects the initially laterally propagating jet. As observed by EUI/HRI, the initially narrow jet reverses its propagation in the direction parallel to the solar limb and starts to expand. This is apparent in both the 174~\AA\ and Lyman-$\alpha$ passbands, with the higher resolution of the 174~\AA\ passband also suggesting a slight untwisting of the jet as it initially erupts. 

This overall behaviour of the jet is also seen in the SDO/AIA passbands shown in Figure~\ref{fig:context}, with the different viewpoint and additional passbands providing additional information. From this viewpoint, the clear nonradial motion is consistent with its interpretation as a ``$\lambda$'' style jet. The jet also appears to be guided radially outward from the Sun following the kinked evolution (although the kinking is less pronounced here), with little to none of the spreading observed by EUI/HRI. This lack of spread in the jet in AIA images could be due to the line-of-sight effects. While the jet can be identified in the 193~\AA\ and 211~\AA\ passbands (observing plasma at $\sim$1--2~MK), it is clearest in the 171~\AA\ and particularly the 304~\AA\ passbands, suggesting that the feature is likely composed of cooler plasma, which is consistent with the clear identification of the jet in the EUI/HRI Lyman-alpha passband.

\corr{
\subsection{Initial evolution} \label{subsec:initial_evol}

\begin{figure*}[!t]
    \centering
    \includegraphics[width=0.87\textwidth]{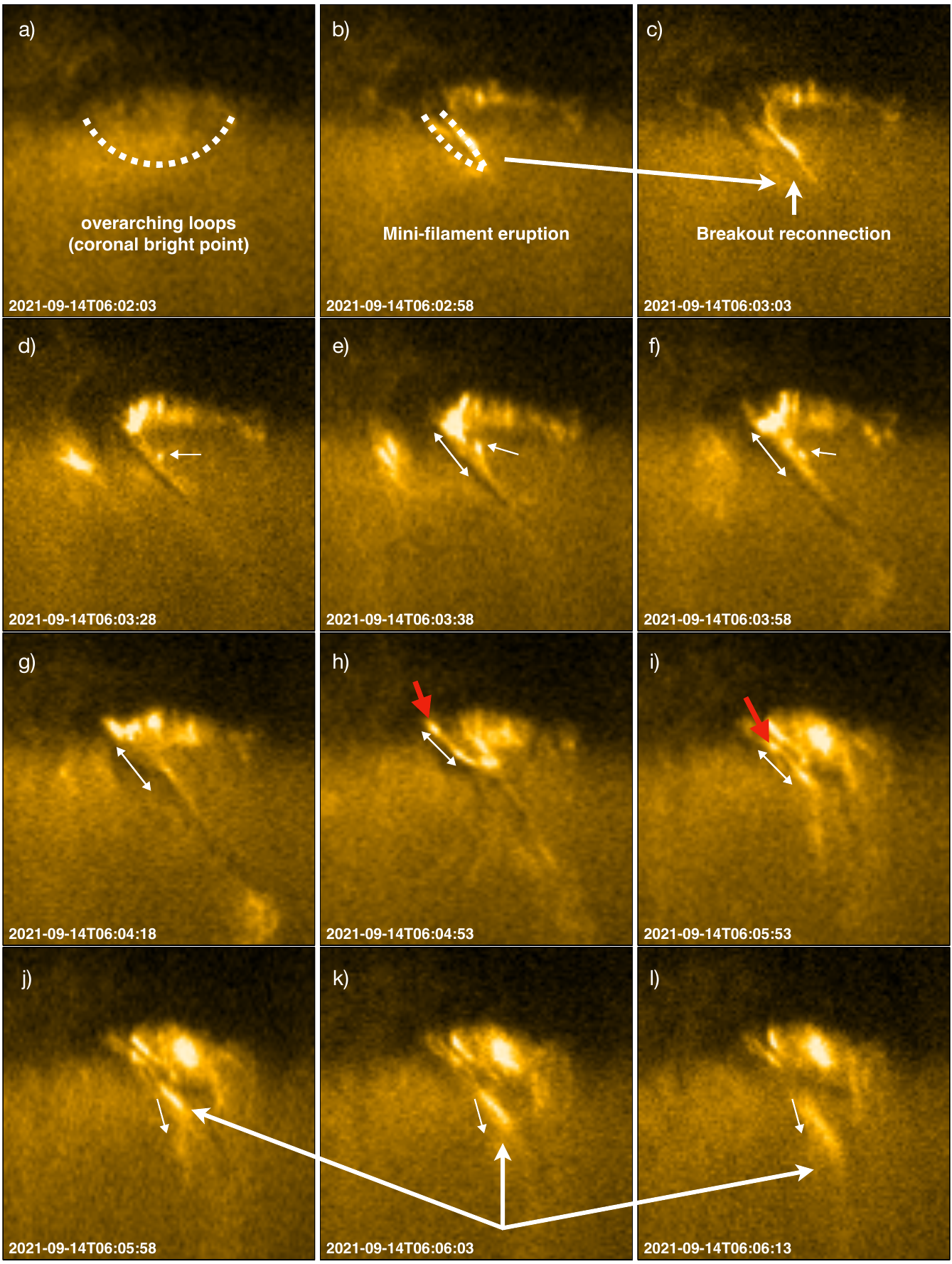}
    \caption{The initial evolution of the jet eruption observed in the 174~\AA\ passband by EUI/HRI. Panel~a shows the initial overarching loops prior to the onset of the eruption. Within $\sim$60~s, (panel~b), the mini-filament starts to erupt, interacting with the overarching loops and driving breakout reconnection (panel~c). Bright plasma blobs produced in this reconnection process then start to drain down the leg of the loops (white arrow in panels~d-f), with the spine of the jet indicated by the double-ended white arrow in panels~e-j. The reconnection opens magnetic field lines, enabling the eruption of the jet plasma (indicated by the red arrow in panels~h-i and subsequently by the white arrow in panels~j-l).}
    \label{fig:eui_evolution}
\end{figure*}

The very high spatial and temporal cadence of the EUI/HRI 174~\AA\ passband enables a detailed analysis of the initial stages of the jet eruption, as shown in Figure~\ref{fig:eui_evolution}. Prior to the onset of the eruption, overarching loops above the coronal bright point can be identified, as shown in Figure~\ref{fig:eui_evolution}a. Within $\sim$60~s, the reconnection process has begun, with the eruption of the mini-filament identifiable in Figure~\ref{fig:eui_evolution}b. This mini-filament then interacts with the overlying loops to drive breakout reconnection as shown in the image 5~s later (Figure~\ref{fig:eui_evolution}c). This can be seen as rapid disconnection and opening of the overarching loops.

The result of bi-directional flows induced by this breakout reconnection can be seen in the evolution of small plasmoid-like blobs draining from the reconnection site down the legs of the overarching loops to the solar surface (as indicated by the unidirectional white arrows in Figure~\ref{fig:eui_evolution}d-e). Meanwhile the mini-filament continues to erupt, with its spine indicated by the bi-directional arrow in panels~e-j of Figure~\ref{fig:eui_evolution}. However, as it erupts, its legs start to stretch (Figure~\ref{fig:eui_evolution}g) and subsequently break, producing bright side lobes (Figure~\ref{fig:eui_evolution}h-i). This enables the eruption of more plasmoid material into the erupting jet (indicated by the red arrow in Figure~\ref{fig:eui_evolution}h \& i and then by the white arrow in Figure~\ref{fig:eui_evolution}j-l) which then evolves out into the solar atmosphere.

In total, this process takes $\sim$140~s from the onset of the eruption to the erupting plasma escaping out into the outer corona, with the very high temporal and spatial resolution of EUI/HRI enabling a detailed analysis of the different reconnection stages of the initiation process. While the same process was also observed by the Lyman-$\alpha$ passband, the lower spatial resolution and very high intensity of the plasma in this passband made it difficult to make a direct comparison of this small-scale behaviour. The difference between the two passbands can be seen in panels~a \& c of Figure~\ref{fig:context}, with the small-scale loop features identifiable in the 174~\AA\ image (panel~c) not as apparent in the Lyman-$\alpha$ image (panel~a).

}

\subsection{Jet kinematics} \label{subsec:kins}

\begin{figure}[!t]
    \centering
    \includegraphics[width=0.49\textwidth]{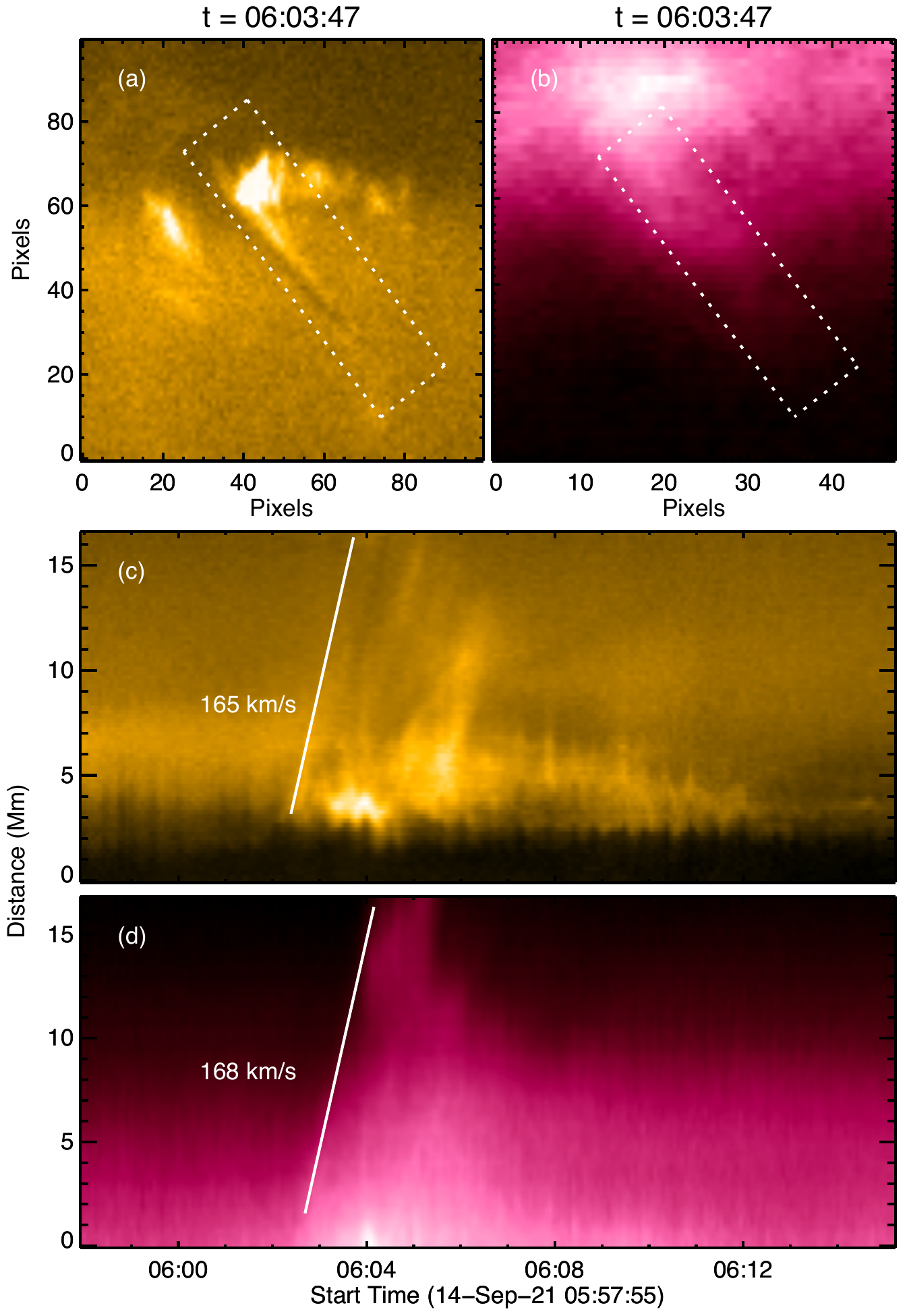}
     \caption{The initial stages of the jet eruption observed in the 174~\AA\ and Lyman-$\alpha$ passbands by EUI/HRI. Panels a \& b show the initial stages of the eruption in the 174~\AA\ and Lyman-$\alpha$ passbands respectively, with the white dotted box in panels a \& b used to produce the distance-time stack plots shown in panels c \& d. A fiducial line is overlaid on the leading edge of the bright jet feature in panels c \& d to guide the eye and to illustrate the propagation speed of the jet.}
    \label{fig:jet_kins_lwr}
\end{figure}

\begin{figure*}[!t]
    \centering
    \includegraphics[width=0.9\textwidth]{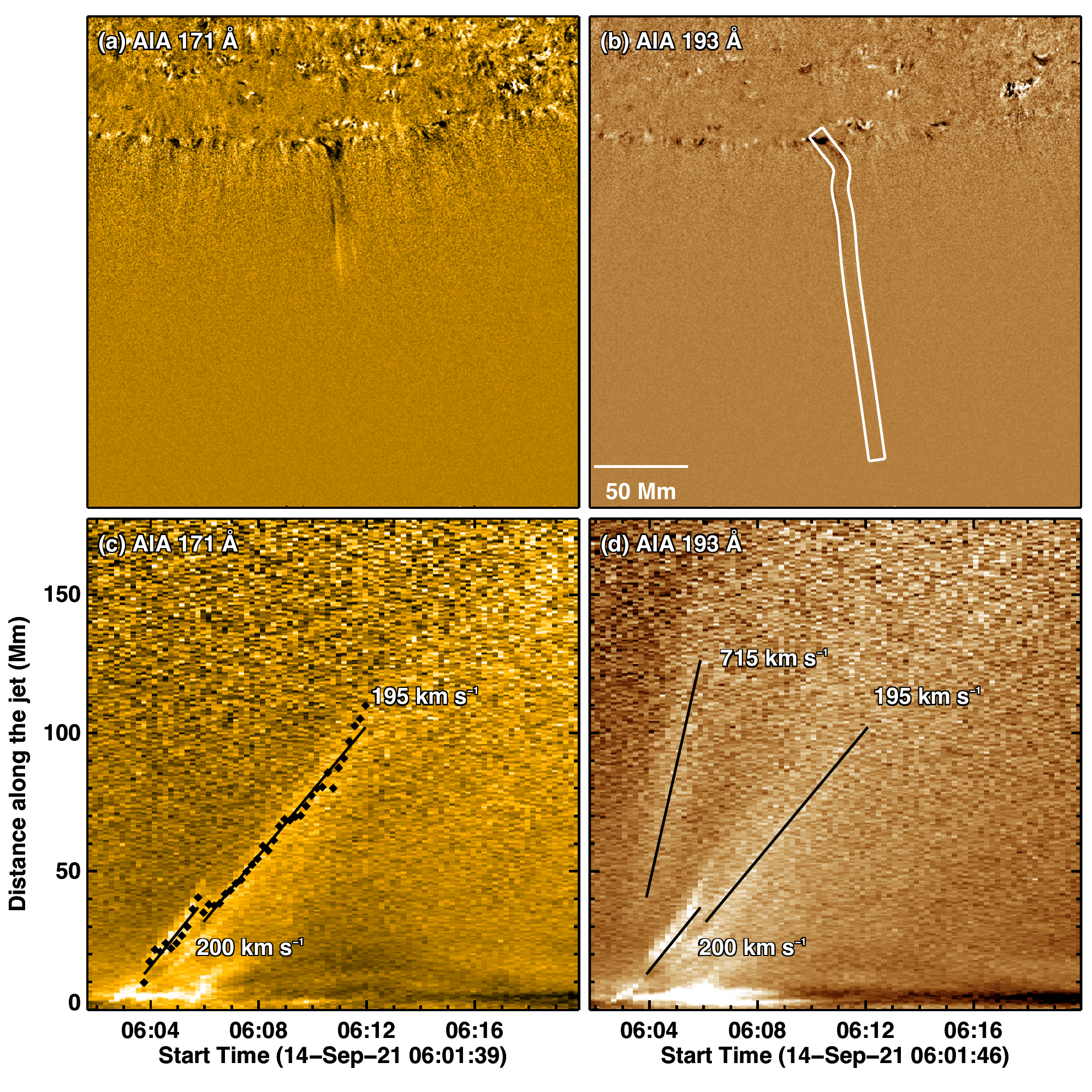}
    \caption{The longer term evolution of the jet eruption as observed using the 171~\AA\ and 193~\AA\ passbands on SDO/AIA. Top row shows the field of view used to track the jet evolution, with the bottom row showing the temporal evolution along the region highlighted in white in panel~b for the 171~\AA\ (left) and 193~\AA\ (right) passbands.}
    \label{fig:aia_jet}
\end{figure*}

\corr{Once the plasma from the jet had been ejected via the breakout reconnection process, the next step was to quantify its evolution.}
A series of distance-time stack plots were used to further examine the temporal evolution of the jet as shown in Figures~\ref{fig:jet_kins_lwr} and \ref{fig:aia_jet} respectively. The initial evolution of the jet close to the origin was examined using observations from EUI/HRI as shown in Figure~\ref{fig:jet_kins_lwr}. The top row of Figure~\ref{fig:jet_kins_lwr} shows individual snapshots of the initial stages of the erupting jet as observed by EUI/HRI 174~\AA\ (panel~a) and EUI/HRI Lyman-$\alpha$ (panel~b) at 06:03:48~UT. Panels c \& d then show the distance-time stack plot along the white box indicated in panels a \& b. The white fiducial lines here indicate that the jet had an initial velocity of $\sim$166~km~s$^{-1}$ as observed by both EUI/HRIEUV and EUI/HRILYA. 


\begin{figure*}[!t]
    \centering
    \includegraphics[width=\textwidth]{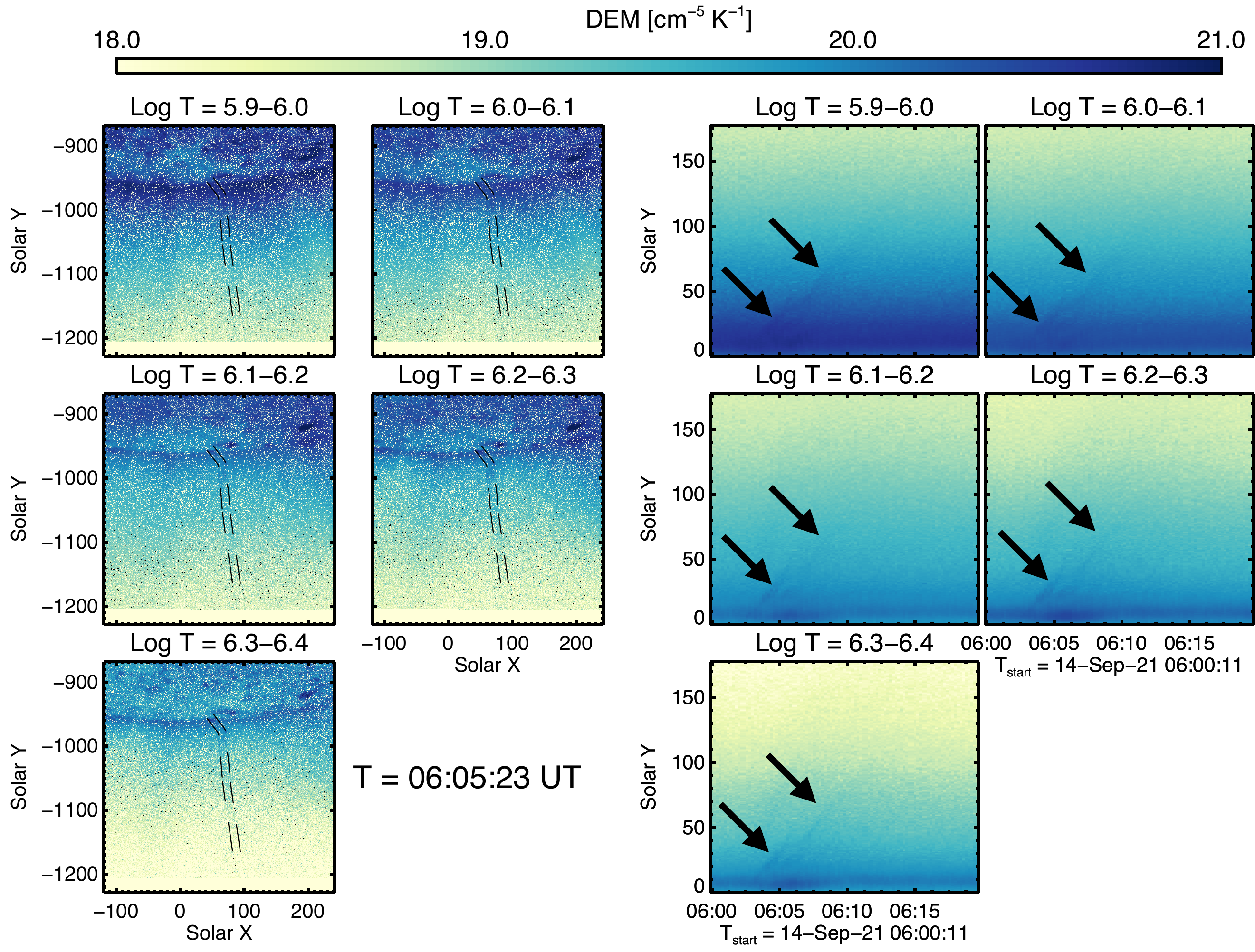}
    \caption{Images (left two columns) and stackplots (right two columns) for different DEM temperature bins (as defined in the panel titles) showing the evolution of the jet in each temperature bin. The propagating jet feature has been highlighted using black arrows in the stackplots to help guide the eye.}
    \label{fig:dem_stack}
\end{figure*}

Following this initial evolution, the longer term evolution of the erupting jet was tracked using the larger field of view of SDO/AIA (see Figure~\ref{fig:aia_jet}). Here, the top row shows the evolution of the erupting jet, with the bottom row showing distance-time stackplots along the white region defined in panel~b for the 171~\AA\ (left column) and 193~\AA\ (right column) passbands. The bright jet feature was then identified in the stackplots as the positions or points of the maximum intensity associated with the jet at each time-step (as shown with symbols in panel~c for the 171~\AA\ passband). These points were then fitted using a linear fit to derive the kinematics. This reveals a slight deceleration in the evolution of the jet as observed by both passbands, from $\sim$200~km~s$^{-1}$ close to the source to $\sim$195~km~s$^{-1}$ further out following the interaction with the nearby faint streamer. 

Although care has been taken to try and ensure that the same feature was identified using the datasets from the different spacecraft, there may be several reasons why the derived kinematics differ slightly. The EUI/HRI observations are taken very close to the source of the erupting jet while the SDO/AIA observations cover a much longer distance and therefore timeframe. The angle of the evolution of the jet with respect to the plane of sky will also affect the kinematics derived using each instrument. 

While both SDO/AIA passbands in Figure~\ref{fig:aia_jet} show a feature propagating outward from the Sun with a fitted velocity of $\sim$200~km~s$^{-1}$, it is also possible to identify a very faint feature in the 193~\AA\ passband with a velocity of $\sim$715~km~s$^{-1}$ propagating ahead of the jet. This feature is not very clear in the SDO/AIA observations presented here, and could not be easily identified in coronagraph images from the LASCO C2 coronagraph onboard the SOHO spacecraft, making it difficult to quantify. However, this feature is consistent with the previous observations by \citet{Cirtain:2007} of two velocities associated with x-ray jets from a polar coronal hole. \citet{Pariat:2015} subsequently suggested that an inclined solar jet could produce a standard jet followed by a helical jet due to the untwisting of newly reconnected magnetic field lines as the eruption evolves. This standard jet could then be observed as a wave travelling at a phase speed close to the local Alfv\'{e}n speed \citep[$\sim$800~km~s$^{-1}$ as measured \corr{in X-ray observations} by][]{Cirtain:2007}, with the helical (or blowout) jet observed as a bulk plasma flow travelling at a fraction of the phase speed. The faint feature observed here \corr{using EUV observations} can therefore be interpreted as the standard jet wave simulated by \citet{Pariat:2015} and observed \corr{in X-rays} by \citet{Cirtain:2007}.

\subsection{Differential Emission Measure analysis} \label{subsect:DEM}

\begin{figure*}[!t]
    \centering
    \includegraphics[width=\textwidth]{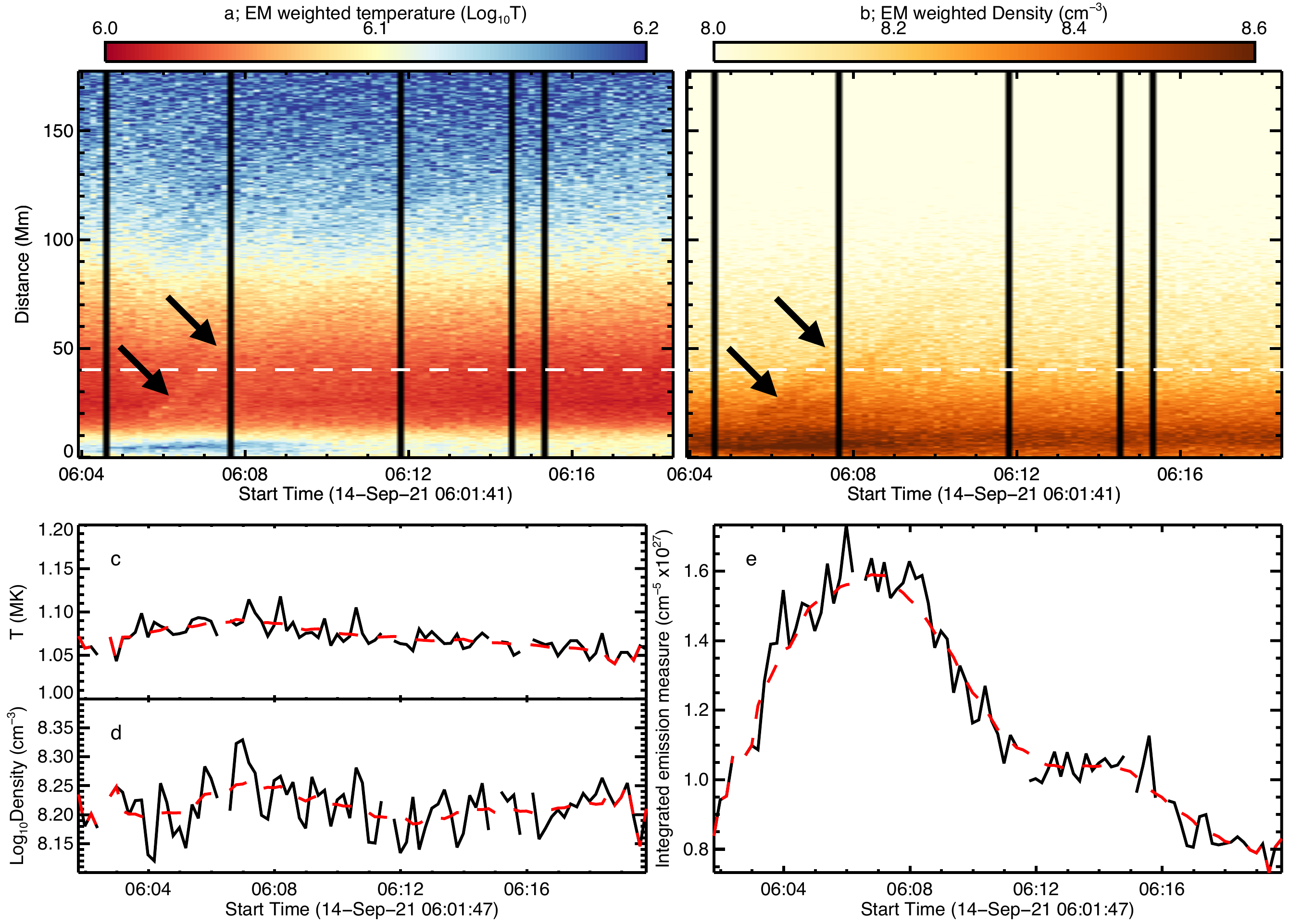}
    \caption{The differential emission measure of the erupting jet. Top row shows the emission measure weighted temperature (left) and electron number density (right) along the region of interest shown in Figure~\ref{fig:aia_jet}b used to calculate the jet kinematics. The faint jet feature has been highlighted using black arrows to help guide the eye. Panels~c \& d show a cut in temperature and the number density respectively along the white dashed line shown in panels a \& b, with the original data shown in black and a heavily smoothed line shown in red to help guide the eye. Panel~e shows the evolution in integrated emission measure in the blue square at the source of the jet eruption shown in Figure~\ref{fig:context}e.}
    \label{fig:dens_temp}
\end{figure*}

The evolution of the jet plasma was examined in more detail using a Differential Emission Measure (DEM) approach in order to quantify how the temperature and density of the jet plasma evolved with time. The DEM $\phi(T)$ is defined as,
\begin{equation}
    \phi(T) = n_e^2(T)\frac{dh}{dT}, \label{eqn:dem}
\end{equation}
where $n_e$ is the electron number density and $h$ is the line of sight depth. Computing DEMs from observations is an ill-posed problem, and multiple techniques have been described to try and solve it using observations from SDO/AIA \citep[see, e.g.,][for more details]{Long:2021}. In this case, we used the Regularised Inversion technique developed by \citet{Hannah:2012,Hannah:2013} to derive the differential emission measure of the field of view shown in Figure~\ref{fig:context}. A selection of the resulting DEMs can be seen in Figure~\ref{fig:dem_stack} for temperatures of $10^{5.9}$ -- $10^{6.4}$~K in bins of $10^{0.1}$~K. Note that above this temperature it was not possible to identify any signature of the erupting jet feature. The left two columns in Figure~\ref{fig:dem_stack} show the DEM images in each temperature bin at t$=06:05:23$~UT, with the right two columns showing the temporal evolution of the DEM in each temperature bin for the region shown in Figure~\ref{fig:aia_jet}b used to calculate the kinematics of the jet eruption as observed by SDO/AIA. Although very faint, a signature of the jet eruption can be identified in the different DEM temperature bins shown here, with the feature identified by arrows in each DEM stackplot to help guide the eye. This faint feature matches the feature observed in Figure~\ref{fig:aia_jet} propagating at a velocity of $\sim$200~km~s$^{-1}$, with no evidence of the faster feature identified in the 193~\AA\ shown in Figure~\ref{fig:aia_jet}d.

To try and further understand the plasma evolution of the jet, the EM-weighted temperature and density of the field of view was estimated using the approach of \citet{Vann:2015,Long:2019}. The EM-weighted electron number density can be defined as,
\begin{equation}
    n_e = \sqrt{\frac{\int\phi(T)dT}{h}}, \label{eqn:dens}
\end{equation}
where $h$ is the plasma scale height, while the DEM-weighted temperature can defined as,
\begin{equation}
    T = \sqrt{\frac{\int\phi(T)TdT}{\int\phi(T)dt}}, \label{eqn:temp}
\end{equation}
\citep[see, e.g.,][]{Cheng:2012}. The resulting temperature and number density stackplots along the jet region highlighted in Figure~\ref{fig:aia_jet}b are shown in panels~a and b of Figure~\ref{fig:dens_temp} respectively. The erupting jet can be identified (and is highlighted using arrows to help guide the eye), but is again very faint in both number density and temperature plots. A profile of both temperature and density along the white dashed line shown in Figure~\ref{fig:dens_temp}a \& b is shown in black in panels c \& d for temperature and number density respectively. In both cases the temporal evolution is very noisy, and a heavily smoothed line has been added in red to highlight the slight increase above the background corresponding to the passage of the jet feature. 

\subsection{Helical Morphology of the Jet and Link to Switchbacks} \label{subsect:kink}

An apparent helical (or kink, twist) morphology of the jet material can be seen in panels~a--d of Figure~\ref{fig:context}. Such morphologies in jets are not uncommon, as noted by e.g., \citet{Shimojo:1996, Wang:1998, Veselovsky:1999, Jiang:2007, Moore:2015}. They suggest the presence of helical or kinked magnetic field lines, similar to structures observed in erupting prominences. We note that the jet is also visible in the Lyman-$\alpha$ HRI channel (Figure~\ref{fig:context}a, b), indicating that a part of the erupting material is at chromospheric or transition region temperatures \citep[see e.g.,][]{Canfield:1996, Sterling:2015}. As noted above, helical field lines in jets are usually interpreted as the result of interchange reconnection of large-scale open coronal hole field lines with small-scale closed field lines at the coronal base \citep{Pariat:2009}. 

The kinked magnetic field lines in jets may be linked to the magnetic field switchbacks that are frequently observed in the near-Sun solar wind by the Parker Solar Probe mission \citep[PSP;][]{Kasper:2019, Bale:2019}. Switchbacks represent significant deviations of the magnetic field from the nominal Parker spiral, which in extreme cases may result in the inversion of the radial component of the magnetic field. The link of coronal jets with switchbacks was suggested by \citet{Sterling:2020} who argued that slightly kinked magnetic field lines resulting from the interchange reconnection may evolve to switchbacks as they propagate in the heliosphere. 

The size of the observed kinked jet in the transverse (i.e. perpendicular to the radial) direction can be estimated from Figure~\ref{fig:context}c to be around 6$\times$10$^3$~km. At this moment, the jet is located at 1.03~$R_\odot$ from the centre of the Sun. Assuming the radial expansion of structures propagating outwards, the transverse size of the corresponding feature at 35.7~$R_\odot$ (the distance of the first perihelion of Parker Solar Probe) would be around 7$\times$10$^6$~km. This is much larger than the transverse scale of switchbacks at this distance from the Sun, which is reported to be around 10$^4$~km \citep{Horbury:2020}. However, Figure~\ref{fig:context}c shows that the jet does not look like a single kinked feature, but rather has a developed fine structure with many individual sub-features at scales down to a few HRI pixels, oriented at different directions with respect to the radial. If a kink has the minimal resolved transverse size of two HRI pixels (i.e. around 400~km at the time of our observations), then its radial expansion would lead to the switchback transverse size of around 5$\times$10$^5$~km at 35.7~$R_\odot$. This size is still an order of magnitude larger than the maximal switchback size (not necessarily in the transverse direction) of 7$\times$10$^4$~km inferred by \citet{Horbury:2020}. This means that the origin of individual switchbacks in the corona cannot be resolved by HRI, even if non-radial structures at the resolved scales of the jet may suggest the presence of helical fields at even smaller scales. The situation will improve only slightly at the closest perihelion of Solar Orbiter around 0.284~au, as the linear spatial resolution of HRI observations will be only a factor two better than that of the observations taken at 0.587~au reported here. 

Another important scale is that of switchback patches, which contain multiple smaller-scale switchbacks \citep{Bale:2021}. Each switchback patch typically corresponds to the supergranulation scales of around 3$^\circ$--5$^\circ$ in heliographic longitude \citep{Bale:2021}, which is a few times larger than the transverse size of our jet (around 0.5$^\circ$). The jet reported in this study therefore corresponds to an intermediate scale situated between the scales of individual switchbacks and the switchback patches.

\section{Discussion \& Conclusions} \label{sec:disc}

As part of its commissioning activities taking observations of the southern polar coronal hole on 2021-Sept-14, Solar Orbiter/EUI observed a small coronal hole jet eruption with very high spatial and temporal resolution. Despite the difference in spacecraft position, the jet was also well observed by SDO/AIA, enabling a multi-viewpoint, multi-thermal analysis of the eruption. In both cases, the jet was observed to initially erupt almost laterally, with a clear kink seen in the jet evolution as it impacted a nearby polar plume observed by both EUI/HRI and SDO/AIA (see movie\_0.mp4 associated with Figure~\ref{fig:context}). The jet was then observed by SDO/AIA to evolve radially away from the Sun, while the jet appeared to reverse its propagation in the direction parallel to the solar limb while propagating outward from the Sun as observed by EUI/HRI.

\corr{The very high spatial and temporal resolution provided by EUI/HRI in the 174~\AA\ passband enabled a detailed analysis of the initiation of the jet. As outlined in Figure~\ref{fig:eui_evolution}, the jet formed as a result of breakout reconnection between the erupting mini-filament and a small overlying loop system. This resulted in the further stretching and reconnection along the erupting filament, with small plasmoid-like blobs visible flowing away from the site of the reconnection down the legs of the loop system. This reconnection opened the overlying magnetic field which then enabled the eruption of the jet itself. Although the process was also observed using the Lyman-$\alpha$ passband, the lower resolution and bright nature of the plasma in this passband meant that it was not possible to discern any fine structure comparable to that observed using the 174~\AA\ passband.}

As observed in Figures~\ref{fig:jet_kins_lwr} and \ref{fig:aia_jet}, the jet displays slightly different kinematics when observed by SDO/AIA and EUI/HRI. Figure~\ref{fig:jet_kins_lwr} shows the initial evolution of the jet observed close to the source prior to the interaction with the nearby polar plume as seen by EUI/HRI 174~\AA\ and Lyman-$\alpha$, with Figure~\ref{fig:aia_jet} showing the longer term evolution of the jet as observed by SDO/AIA. These figures suggest a constant linear velocity for the jet of $\sim$165-200~km~s$^{-1}$, with the slight differences between the observations from the two spacecraft most likely due to the different angle of observation, with the jet exhibiting different kinematics when propagating out of the plane of sky as seen from different perspectives. The jet was also best observed using the cooler passbands available from both SDO/AIA and EUI/HRI, with a very clear jet seen in both AIA 304~\AA\ and HRI Lyman-$\alpha$, although it was possible to discern a fainter feature in the 171/174~\AA, 193~\AA, and 211~\AA\ passbands. A very faint feature was also observed in the SDO/AIA 193~\AA\ passband propagating away from the origin with a velocity of $\sim$715~km~s$^{-1}$ as shown in Figure~\ref{fig:aia_jet}. This is consistent with the previous observations \corr{in X-rays} by \citet{Cirtain:2007} and simulations of \citet{Pariat:2015} of a wave travelling at a phase velocity close to the Alfv\'{e}n speed ahead of bulk plasma motion in a coronal jet, and implies that the jet is the result of interchange reconnection between open and closed magnetic field in the corona, consistent both with the blowout jet interpretation, and the \corr{unique high resolution} EUI/HRI observations \corr{in EUV of the breakout reconnection in the initial stages of the eruption, and the observations of} an initial untwisting of the jet as it erupted. Although this behaviour could be linked to the switchbacks detected by PSP, the size of the kink observed here is much larger than the typical switchback size measured by \citet{Horbury:2020}.

While the jet can be clearly seen in the individual passbands shown in Figure~\ref{fig:context}, it was much more difficult to identify using the derived differential emission measure, as can be seen in Figure~\ref{fig:dem_stack}. The jet can be identified using the temperature bins shown in the figure, but no signal was observed in the temperature bins above $\sim$10$^{6.4}$~K. A further analysis of the emission measure weighted temperature and number density (as shown in Figure~\ref{fig:dens_temp}) show that a slight increase in both temperature and density could be identified associated with the passage of the jet, but it was very small in both cases. These observations, combined with the very clear presentation of the jet in the 304~\AA\ passband observed by SDO/AIA and the Lyman-$\alpha$ observed by EUI/HRI, indicate that the jet primarily consisted of cooler material which was below the threshold of the DEM calculated using SDO/AIA observations. This indicates that the jet was erupting chromospheric material, \corr{again} consistent with a blowout jet interpretation.

However, the DEM derived from SDO/AIA can also be used to estimate the radiative thermal energy associated with the onset of the jet eruption. As discussed by \citet{Benz:1999} and more recently by \citet{Purkhart:2022}, it is possible to estimate the radiative thermal energy $E_{th}$ released during a coronal brightening using the evolution of emission measure via the equation,
\begin{equation}
    E_{th} = 3 k_B T \sqrt{\Delta EM q h A}, \label{eqn:energy}
\end{equation}
where $k_B$ is Boltzmann's constant, $T$ is the temperature during the peak of the EM evolution, $\Delta EM$ is the change in integrated emission measure relative to the pre-event value measured in cm$^{-5}$, $h$ is the total line of sight thickness of the event, $A$ is the total area of the event, and $q$ is the filling factor, representing the difference between observed and actual volumes occupied by the emitting plasma. As this is difficult to accurately estimate, and does not measurably affect the estimated energy, we have assumed a filling factor $q$ of 1 \citep[cf.][]{Parnell:2000,Purkhart:2022}. The region chosen here as corresponding to the onset of the jet eruption is indicated by the blue square in Figure~\ref{fig:context}e, with the temporal evolution of the integrated emission measure in this region shown in Figure~\ref{fig:dens_temp}e. Note that due to the highly variable nature of the evolution of the emission measure, the $\Delta EM$ was defined using the change between the start and peak of the smoothed red dashed line shown in Figure~\ref{fig:dens_temp}e. We also followed the lead of \citet{Purkhart:2022}, assuming a line of sight distance $h = \sqrt{A}$, after \citet{Benz:1999}. From equation~\ref{eqn:energy}, we derived a radiative thermal energy for the source region of the coronal jet of $1.5\times10^{24}$~ergs, comparable to the energy of a nanoflare \citep[see e.g.][]{Chitta:2021}. However, it is worth noting that this is most likely an underestimation of the true radiative thermal energy produced by the source region\corr{. The jet in this case erupted from very close to the limb, but still on-disk as observed by SDO/AIA, and this} close proximity to the limb \corr{leads to an} increased absorption of the EUV radiation along the line of sight to the observer.


These observations highlight the importance of high time cadence observations with very high spatial resolution when studying the evolution of small scale features in the solar atmosphere. The HRI/EUV observations of the erupting jet presented here \corr{show the different stages of the breakout reconnection process which led to the eruption of the jet in very fine detail, and} reveal untwisting of the jet as it erupts, while the Lyman-$\alpha$ observations show the cool nature of the erupting plasma. The additional point-of-view of Solar Orbiter is also very useful, as it highlights the interaction of the erupting jet with the nearby polar plume; behaviour that would have been missed if only observations from SDO/AIA were used to study the jet. However, the multiple passbands provided by SDO/AIA are vital for understanding the plasma diagnostics associated with the erupting jet, and for estimating the energy released during the eruption. The combination of observations from both SDO and Solar Orbiter will offer new insights into coronal jets, particularly as Solar Orbiter moves out of the solar ecliptic towards the poles.

\begin{acknowledgments}
 The authors wish to thank the anonymous referee whose suggestions helped to improve the paper. DML wishes to thank Peter Wyper and Etienne Pariat for useful discussions which helped to codify the interpretation of the jet presented in this paper. Solar Orbiter is a space mission of international collaboration between ESA and NASA, operated by ESA. The EUI instrument was built by CSL, IAS, MPS, MSSL/UCL, PMOD/WRC, ROB, LCF/IO with funding from the Belgian Federal Science Policy Office (BELSPO/PRODEX PEA 4000134088); the Centre National d’Etudes Spatiales (CNES); the UK Space Agency (UKSA); the Bundesministerium f\"{u}r Wirtschaft und Energie (BMWi) through the Deutsches Zentrum f\"{u}r Luft- und Raumfahrt (DLR); and the Swiss Space Office (SSO). SDO data are courtesy of NASA/SDO and the AIA, EVE, and HMI science teams. D.M.L. is grateful to the Science Technology and Facilities Council for the award of an Ernest Rutherford Fellowship (ST/R003246/1). L.P.C. gratefully acknowledges funding by the European Union. Views and opinions expressed are however those of the author(s) only and do not necessarily reflect those of the European Union or the European Research Council (grant agreement No 101039844). Neither the European Union nor the granting authority can be held responsible for them. D.Baker and I.G.H are funded under STFC consolidated grant numbers ST/S000240/1 and ST/T000422/1 respectively. N.N is supported by STFC PhD studentship grant ST/W507891/1. A.N.Z. thanks the Belgian Federal Science Policy Office (BELSPO) for the provision of financial support in the framework of the PRODEX Programme of the European Space Agency (ESA) under contract number 4000136424.
\end{acknowledgments}

\facilities{SDO/AIA, Solar Orbiter EUI}

\software{SSW/IDL \citep{Freeland:1998}}

\bibliography{bibliography}{}
\bibliographystyle{aasjournal}

\end{document}